\documentclass[journal]{IEEEtran}

\usepackage{graphicx,cite,epsfig,amssymb,amsmath}
\usepackage{pifont}
\usepackage{stmaryrd}
\usepackage{bbding}
\usepackage{algorithm}
\usepackage{algorithmic}
\usepackage{mathrsfs}
\usepackage{latexsym}
\usepackage{indentfirst}
\usepackage{fmtcount}
\usepackage{cite}
\usepackage{float}
\usepackage{morefloats}

\usepackage{caption}
\usepackage{subcaption}

\usepackage{color}
\usepackage{mathtools}
\ifCLASSINFOpdf
\else
\fi
\begin{document}
%
\title{Delay-aware Fountain Codes for Video Streaming with Optimal Sampling Strategy}
%
%
%

\author{Kairan~Sun,~\IEEEmembership{Student Member,~IEEE,}
        Huazi~Zhang,~\IEEEmembership{Member,~IEEE,}
        and~Dapeng~Wu,~\IEEEmembership{Fellow,~IEEE}
        }
\maketitle

\begin{abstract}

The explosive demand of on-line video from smart mobile devices poses unprecedented challenges to delivering high quality of experience (QoE) over wireless networks. Streaming high-definition video with low delay is difficult mainly due to (i) the stochastic nature of wireless channels and (ii) the fluctuating videos bit rate. To address this, we propose a novel delay-aware fountain coding (DAF) technique that integrates channel coding and video coding. 
In this paper, we reveal that the fluctuation of video bit rate can also be exploited to further improve fountain codes for wireless video streaming. Specifically, we develop two coding techniques: the time-based sliding window and the optimal window-wise sampling strategy. By adaptively selecting the window length and optimally adjusting the sampling pattern according to the ongoing video bit rate, the proposed schemes deliver significantly higher video quality than existing schemes, with low delay and constant data rate. To validate our design, we implement the protocols of DAF, DAF-L (a low-complexity version) and the existing delay-aware video streaming schemes by streaming H.264/AVC standard videos over an 802.11b network on CORE emulation platform. The results show that the decoding ratio of our scheme is 15\% to 100\% higher than the state of the art techniques.
\end{abstract}


%
\IEEEpeerreviewmaketitle

\section{Introduction}
\label{sec:intro}

In the recent decade, we have witnessed the bloom of video services over the Internet. Some of them provide pre-recorded video streams such as YouTube and Netflix; others provide live video communications such as Skype and Facetime. 
As expected, a huge amount of multimedia contents will be generated and consumed. On the other hand, the prevalent smart mobile devices make these contents more accessible to people than ever. Thanks to the evolution of communication technologies, such as 3G/4G, LTE, WiFi, etc., wireless networks are widely available in our daily lives. However, despite the promising developments, the stochastic nature of wireless channels still persists: its vulnerability to channel noise, inter-user interference and low data rate under mobility. The problems easily deteriorate in video-dominant applications where the requirement on channel quality is the highest. As a result, how to stream videos with low delay, stable data rate and high quality over wireless network raises formidable challenges in communication society.

To provide reliable video transmission, advanced coding and signal processing techniques, such as forward error correction (FEC) erasure codes, have been proposed.
One important class of FEC codes are fountain codes \cite{fountain}, such as Luby transform (LT) code \cite{LTCodes} and Raptor code \cite{RaptorCodes}. Fountain codes are ideal for wireless video streaming for its ratelessness: fountain codes will reconstruct the original data using the redundancy sent by the sender, without demanding acknowledgments (ACK) or retransmissions. 
%
%
%
The traditional fountain codes are initially designed for achieving the \textit{complete} decoding of the \textit{entire} original file. 
That means, if a video file is transmitted using traditional fountain codes, users cannot watch it until the whole video file is successfully decoded. However, a lot of video streaming applications are \textit{delay-aware} and \textit{loss-tolerant}, which means (i) the time interval between video being generated and being played can not exceed a certain threshold; and (ii) partial decoding is tolerable, albeit higher decoding ratio is still preferred. 

In order to introduce delay awareness into fountain codes, the most intuitive solution is to partition the video file into fixed-length data blocks, separately encode them, and transmit sequentially. 
We call this method the \textit{block coding} scheme. From the perspective of video transmission, a smaller block size is preferred, because it leads to shorter playback latency. From the perspective of fountain codes, however, the block size needs to be as big as possible to maintain a smaller coding overhead \cite{overhead}. The fundamental trade-off between video watching experience and coding performance is crucial for the design of delay-aware fountain codes. 

%

In this paper, we propose a novel delay-aware fountain code scheme for video streaming that deeply integrates channel coding and video coding. Our scheme is based on \textit{sliding window fountain codes} (SWFC) \cite{FirstSliding}, which partitions a file into many overlapping data windows and transmits them sequentially. Although there have been a number of works on joint fountain- and video-coding design, such as \cite{FirstSliding,SlidingRaptorULP,UEPFountain,ExpandWindow,ExpandWindow2}, they do not fully exploit the characteristics of multimedia content in order to optimize video watching experience.

The novelty of our scheme lies in that we do not treat the sliding windows as homogeneous, which, according to existing methods, have fixed length and uniform sampling distribution. On the contrary, our scheme optimizes every window and thus is a deep-level joint design of multimedia streaming and channel coding. Innovatively, we take into account video bit rate fluctuation and video coding parameters, such as group of pictures (GOP) size and frame rate, at the level of channel coding, and exploits them in the design of a novel delay-aware fountain coding approach. As a result, the proposed schemes take advantages of all the benefits of fountain codes, and optimize them in the context of delay-aware video applications. According to a novel performance metric, i.e., \textit{in-time decoding ratio}, which better reflects the real video watching experience, our methods significantly outperform existing schemes.

The contributions and key techniques of this paper are summarized below.

\begin{enumerate}

\item We propose a \textit{time-based sliding window scheme} to provide the much desired delay awareness in video streaming. Unlike the existing SWFC schemes that have a fixed number of packets in each window, our scheme adaptively selects window lengths according to the number of bits in frames. In this way, we can maximize the code word length in the coding blocks, so as to achieve higher coding gain within a bounded playback delay.

\item We propose an \textit{optimal window-wise sampling strategy}, in order to deliver a consistent watching experience. Since all the existing SWFCs uniformly sample and encode the packets within each window, due to video bit rate fluctuation, the received video quality may be time-varying. By optimally adjusting the sampling pattern according to the ongoing video bit rate, the proposed technique delivers significantly higher decoding ratio than existing schemes.

\item We develop a \textit{Delay-Aware Fountain codes protocol} (\textit{DAF}) by integrating all the above mentioned techniques to deliver the optimal solution. To reduce computational complexity, we also propose a sub-optimal yet low-complexity version called \textit{DAF-L}. By comparing with its conventional counterparts, such as \cite{FirstSliding,ExpandWindow,ExpandWindow2}, in various scenarios, our approach is shown to yield the best overall performance.

\end{enumerate}

The paper is organized as follows. Section \ref{sec:delay-aware} proposes the time-based sliding window scheme, and compares our distinct designs with related work. Section \ref{sec:sample-dist} proposes our optimal window-wise sampling strategy. 
Section \ref{sec:system} designs the DAF and DAF-L system using all the techniques we proposed in previous sections. Section \ref{sec:experiments} gives the simulation results. Section \ref{sec:conclusions} concludes the paper.

\section{Delay-aware Sliding Window Fountain Codes}
\label{sec:delay-aware}


Our work focuses on a deep integration of fountain codes and video coding, hence the concepts in both fountain codes and video coding will be frequently referred to. 
We define two sets of variables in Table \ref{tab:fountain-define} and \ref{tab:video-define}. Table \ref{tab:fountain-define} lists the variables related to fountain codes, and Table \ref{tab:video-define} lists the properties related to video coding.  


\begin{table}[] 
	\centering
	\caption{Definitions of the notations for the variables related to fountain codes.}
	\begin{tabular}{ l|l|p{120pt} }
		\hline
		\hline
		Notation & Unit & Definition  \\
		\hline
		$P$ & 		byte & 			Packet size. The number of bytes in the payload. \\
		$R$ & 		byte/second & 	Data rate. \\
		$C$ &		N/A & 			Code rate. \\
		$PLR$ &		N/A & 			Packet loss rate.  \\
		$\Delta t$&	frame & 		Step size. The number of frames the window shifts each time it slides forward. \\
		$W$ & 		frame & 		Sliding window size. \\
		$w_{W}(t)$&	packet & 		Number of native packets in the sliding window starting from $t^{\text{th}}$ frame.  \\
		$k$ & 		packet & 		Total number of native packets. \\
		$N$ & 		packet & 		Total number of coded packets. \\
		$N_{W}$ & 	packet & 		Number of coded packets to be sent within each sliding window. \\ 
		$N_{window}$ & window & 	Number of windows. \\
		\hline
		\hline
	\end{tabular}
	\label{tab:fountain-define}
\end{table}%

For notation simplicity, all the concepts relating to ``time'' in this article are actually in the unit of ``number of frames''. 
We can then get the total number of native packets $k$ as $k = \sum\limits_{t = 1}^{T} s(t)$. The definitions of the other variables will be introduced when they are used later in this article.

\begin{table}[] 
	\centering
	\caption{Definitions of the notations for the variables related to video coding. }
	\begin{tabular}{ l|l|p{120pt} }
		\hline
		\hline
		Notation & Unit & Definition \\
		\hline
		$F$ & 		frame/second & 	Frame rate. \\
		$T_{Delay}$ & frame & 		Tolerable end-to-end delay.  \\
		$s(frmno)$ & packet & 		Number of native packets in the $frmno^{\text{th}}$ frame. \\
		$N_{GOP}$ & frame & 		GOP size. The number of frames in a GOP.  \\
		$T$ & 		frame &		 	Video length. Total number of frames in the video sequence. \\
		$pkt(t_0,k) $ & packet&		The number of packets in the first $k$ frames in the window starting from the $t_0^{\text{th}}$ frame of the video. \\
		$pktno(frmno)$ & packet no.&	Starting packet sequence number of the $frmno^{\text{th}}$ frame in the video.  \\
		$frmno(pktno)$ & frame no. &	Frame sequence number from which the $pktno^{\text{th}}$ original packet belongs to. \\ 
		\hline
		\hline
	\end{tabular}
	\label{tab:video-define}
\end{table}%

In the following two sections, we will discuss two key novel designs of the proposed scheme, as opposed to those existing designs. 

\subsection{Sliding Window vs. Block Coding}
\label{sec:sliding}

\begin{figure}[t]
\centering
	\begin{subfigure}[b]{0.35\textwidth}
		\includegraphics[width=1\textwidth]{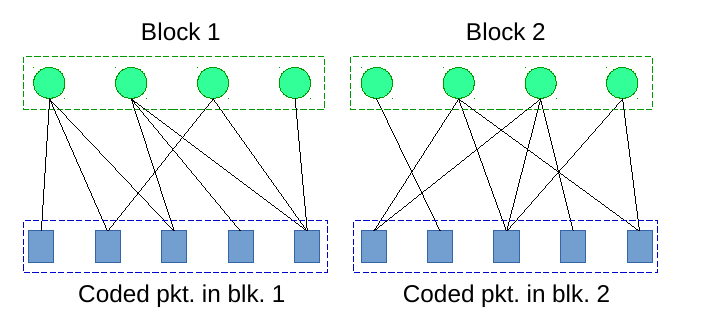}
		\caption{Block coding scheme.}
		\label{fig:blockDecode}
	\end{subfigure}
	\hfil
	\begin{subfigure}[b]{0.35\textwidth}
		\includegraphics[width=1\textwidth]{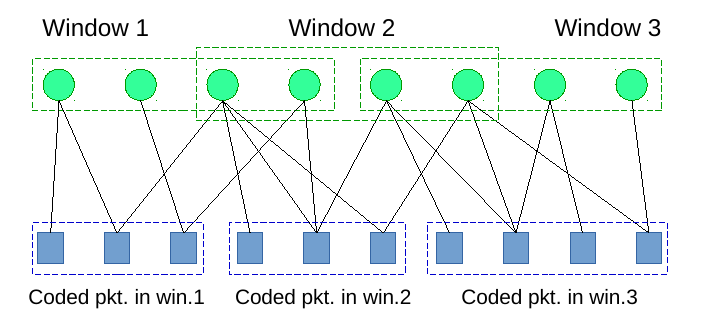}
		\caption{Sliding window scheme.}
		\label{fig:slidingDecode}
	\end{subfigure}
\caption{Comparison of coding structure between (a) block coding scheme, and (b) sliding window scheme. }
\label{fig:blockvsslidingDecode}
\end{figure}

The concept of SWFC was first proposed in \cite{FirstSliding} for LT codes. Then, the similar idea was extended to Raptor codes and unequal error protection (UEP) was applied in \cite{SlidingRaptorULP}. As shown in Fig.~\ref{fig:blockDecode}, the block coding scheme has a relatively small block size, and the coded packets for each block are only linked to the source packets in a small window. But in Fig.~\ref{fig:slidingDecode}, the overlap between sliding windows makes decoded packets in one window to help the decoding of other windows. In that sense, the size of the window is virtually extended. As a result, sliding window schemes virtually extend the block size, so as to enhance the performance of the fountain codes by reducing the overhead.

In \cite{ExpandWindow,ExpandWindow2}, the authors proposed an expanding window fountain code scheme. Instead of using the overlapping fixed-size windows, the packets in each window must be a subset of the next window. Although this scheme is also delay-aware, it is not suitable for video streaming, since the decoding probability is unbalanced: the probability of decoding the frames in the beginning is higher than the latter ones. On the other hand, \cite{UEPFountain} uses a block coding scheme, and its virtual block size is expanded by duplicating all symbols in each block. However, all the above schemes does not carefully examine of the relationship between block size and the end-to-end delay in delay-aware applications.


In our design, we use an SWFC scheme. The step size between two consecutive windows is $\Delta t$. For simplicity, we assume that $W$ and $T$ are integral multiples of $\Delta t$. In order to avoid dividing the frames from one GOP into different windows, $\Delta t$ should be an integral multiple of $N_{GOP}$. 

\begin{figure*}[t]
\centering
	\begin{subfigure}[b]{0.35\textwidth}
		\includegraphics[width=1\textwidth]{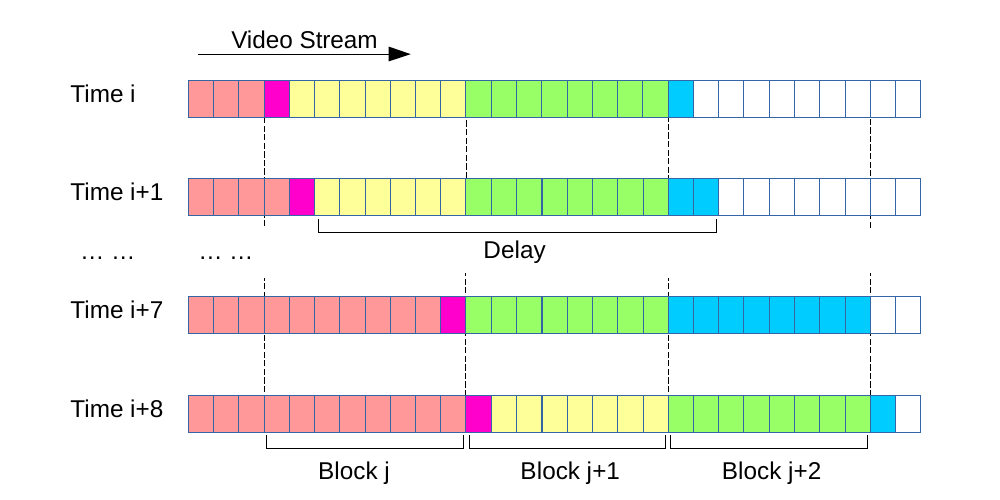}
		\caption{Block coding.}
		\label{fig:block}
	\end{subfigure}
	\hfil
	\begin{subfigure}[b]{0.35\textwidth}
		\includegraphics[width=1\textwidth]{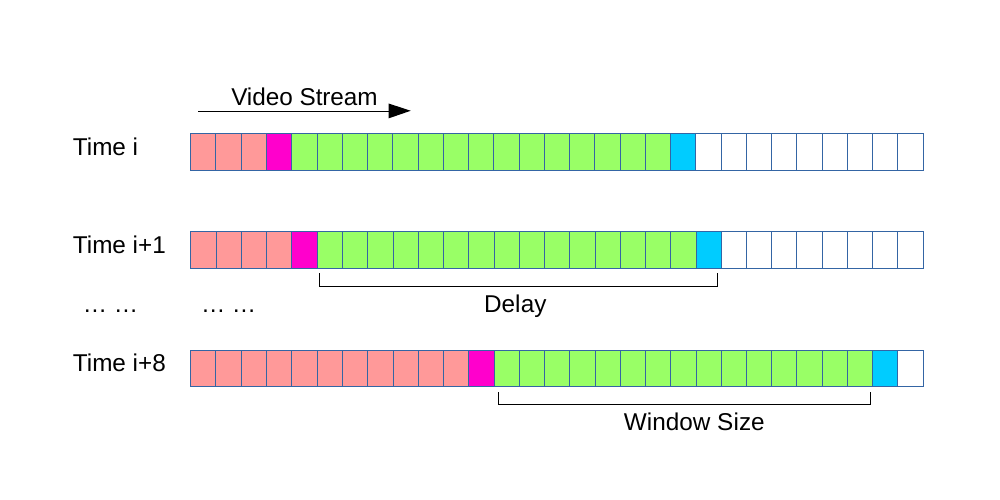}
		\caption{Sliding window when $\Delta t = 1$.}
		\label{fig:sliding}
	\end{subfigure}
	\hfil
	\begin{subfigure}[b]{0.20\textwidth}
		\includegraphics[width=1\textwidth]{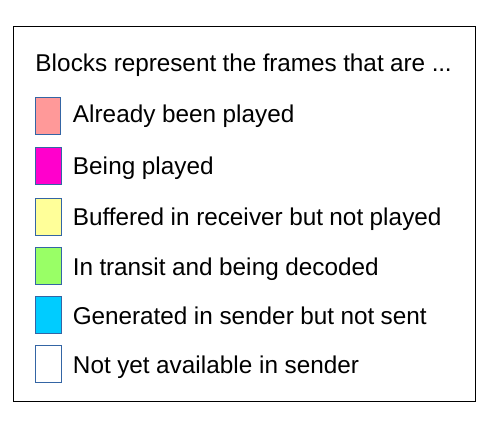}
		\caption{Meanings of the colors in the blocks.}
		\label{fig:legend}
	\end{subfigure}
\caption{Compare block vs. sliding window. }
\label{fig:blockvssliding}
\end{figure*}

One noteworthy difference between SWFC and block coding is the relationship among $\Delta t$, $T_{Delay}$ and window size $W$.
For block coding, as shown in Fig.~\ref{fig:block}, because the receiver can only start to play the content in current block when the transmission for this block finishes, and the sender can only start to encode and send the next block's packets when all the packets in the next block are available, the end-to-end play delay $T_{Delay} \ge 2W$. For SWFC, if the step size is $\Delta t $, the encoder can start to encode the next window as soon as next $\Delta t $ packets are available, so the end-to-end play delay $T_{Delay} \ge W + \Delta t$. The above relationships implicitly impose the maximal window size (which corresponds to the best coding efficiency) we can set for both schemes. If we deem block coding as a special case of sliding window when $\Delta t = W$, we can see the sliding window cannot exceed $\frac{1}{2}T_{Delay}$. We also know that the biggest window size is obtained when $\Delta t = 1$, as shown in Fig.~\ref{fig:sliding}.

The number of windows to be sent, $N_{window}$, could be obtained by $N_{window} = \frac{T-W}{\Delta t}$. A very important derived parameter is $N_{W}$, the number of coded packets to be sent within each sliding window, which is a link between fountain codes and video coding: $N_{W} = \frac{R \cdot \Delta t}{F \cdot P} $. Then, the total number of coded packets to be sent, $N$, can be defined based on $N_{W}$: $N = \lfloor N_{window} \times N_{W} \rfloor = \lfloor  \frac{R \cdot (T - W)}{F \cdot P} \rfloor $. The overall code rate $C$ is then defined using $C=\frac{k}{N}$.


\subsection{Time-based Window Size vs. Packet-based Window Size }
\label{sec:time-based}

The distinct aspect about our proposed scheme is that the size of windows are based on time (or interchangeably speaking, based on the number of frames).

Although the specific methods may vary, a lot of existing work, such as \cite{FirstSliding,SlidingRaptorULP,UEPFountain,ExpandWindow,ExpandWindow2}, designed the delay-aware fountain codes based on the following core idea: group the video data into windows (either overlapping or non-overlapping), and send the windows one by one within each period of time. In the aforementioned work, the coding parameters, such as the size of the windows, the speed of window movement, and the total length of the data, are constant numbers based on the number of packets. Inherently, the authors considered the number of packets as an abstraction of the number of frames. 

However, the packet-based schemes ignored an important characteristic of video data: 
there are different amounts of bits for each frame. Even if rate control techniques are used, they may inevitably lead to bit rate fluctuation and video quality degradation \cite{prd}. This fact makes \textit{packet-based windows different from time-based windows.} Dividing the video streaming data into blocks with fixed number of packets will result in the following phenomenons:

\begin{enumerate}
\item \textit{Improper partition of frames and GOPs}:
Because a frame may contain various number of packets, it is highly likely that packets from one frame, or one GOP, to fall into two blocks, thus causing video playback error. 

\item \textit{Uncontrollable delay}:
Because there are different amounts of frames in each block, the time of delay varies from time to time, so we do not know the resulting time delay. As a result, packet-based window cannot be used in real-time or delay-aware systems.
Even if we have to use packet-based window in delay-aware systems, we need to know the number of packets in each frame of the video before-hand, and select the fewest number of packets in any $T_{Delay}$-frame period as the window size. 
In that case, the tolerable delay $T_{Delay}$ is underused for most of the time, which contradicts the designing principle of making the best use of delay. 

\item \textit{Unstable data rate}:
Using fixed code rate, the encoder will generate the same amount of coded packets within any packet-based window. However, because of the nonuniformity of video bit rate, the data rate will be different in different time periods. 
\end{enumerate}

On the contrary, by using the time-based window size, all the above issues will be resolved: it will ensure one frame to be grouped in a same window; it will make the best use of the delay at all time; if the encoder generates the same amount of coded packets within any time-based window, the data rate will be a constant.




\section{Optimal Window-wise Sampling Strategy}
\label{sec:sample-dist}

\subsection{Nonuniform Global Sampling Distribution Using SWFC}

For an LT encoder, each coded packet is generated using the following two steps: (i) Randomly choose the degree $d_n$ of the packet from a degree distribution $r(d)$; (ii) Choose $d_n$ input packets at random from the original packets uniformly, and a coded packet is obtained as the exclusive-or (XOR) of those $d_n$ packets. 
The optimization of degree distribution $r(d)$ for LT code has been well studied in some literatures, e.g. \cite{LTCodes,optimizeLTImportance,smallMsgDegree}. It should be noted that, the optimization procedures are all built on a common prerequisite -- the sampling distribution should be uniform, because the highest efficiency of fountain codes is achieved using uniform distribution.

\begin{figure}[ht]
\centering
	\begin{subfigure}[b]{0.35\textwidth}
		\includegraphics[width=1\textwidth]{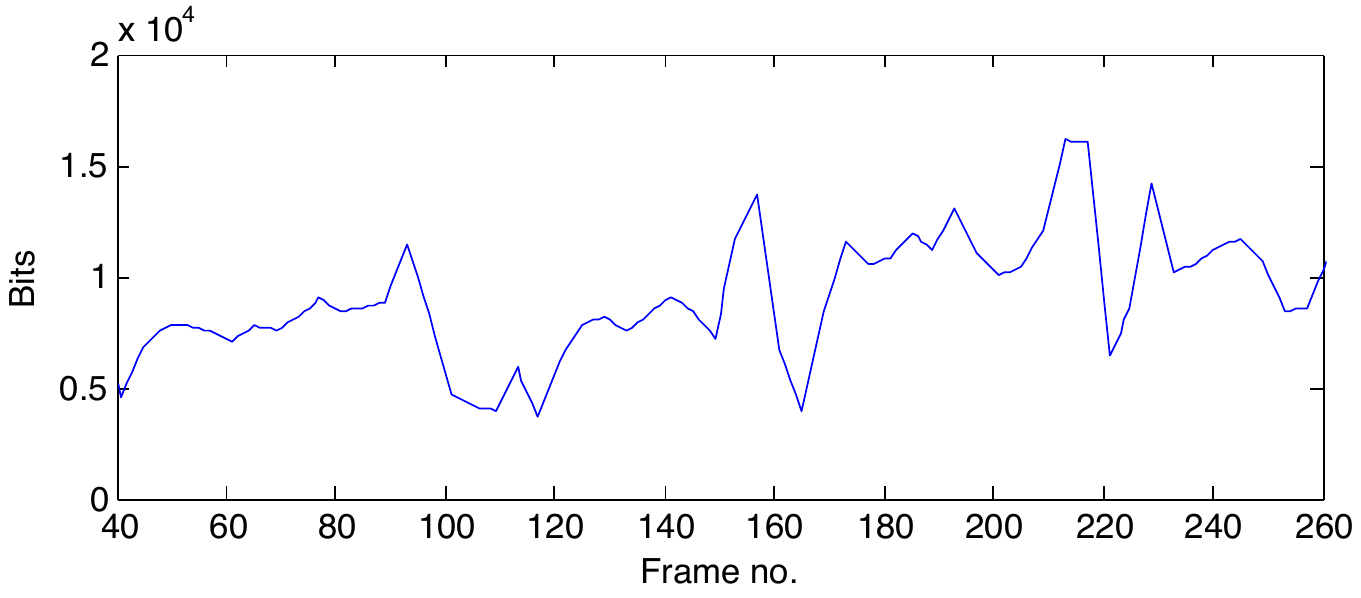}
		\caption{Number of bits in each frame for CIF sequence \textit{foreman}.}
		\label{fig:bitrate}
	\end{subfigure}
	\begin{subfigure}[b]{0.35\textwidth}
		\includegraphics[width=1\textwidth]{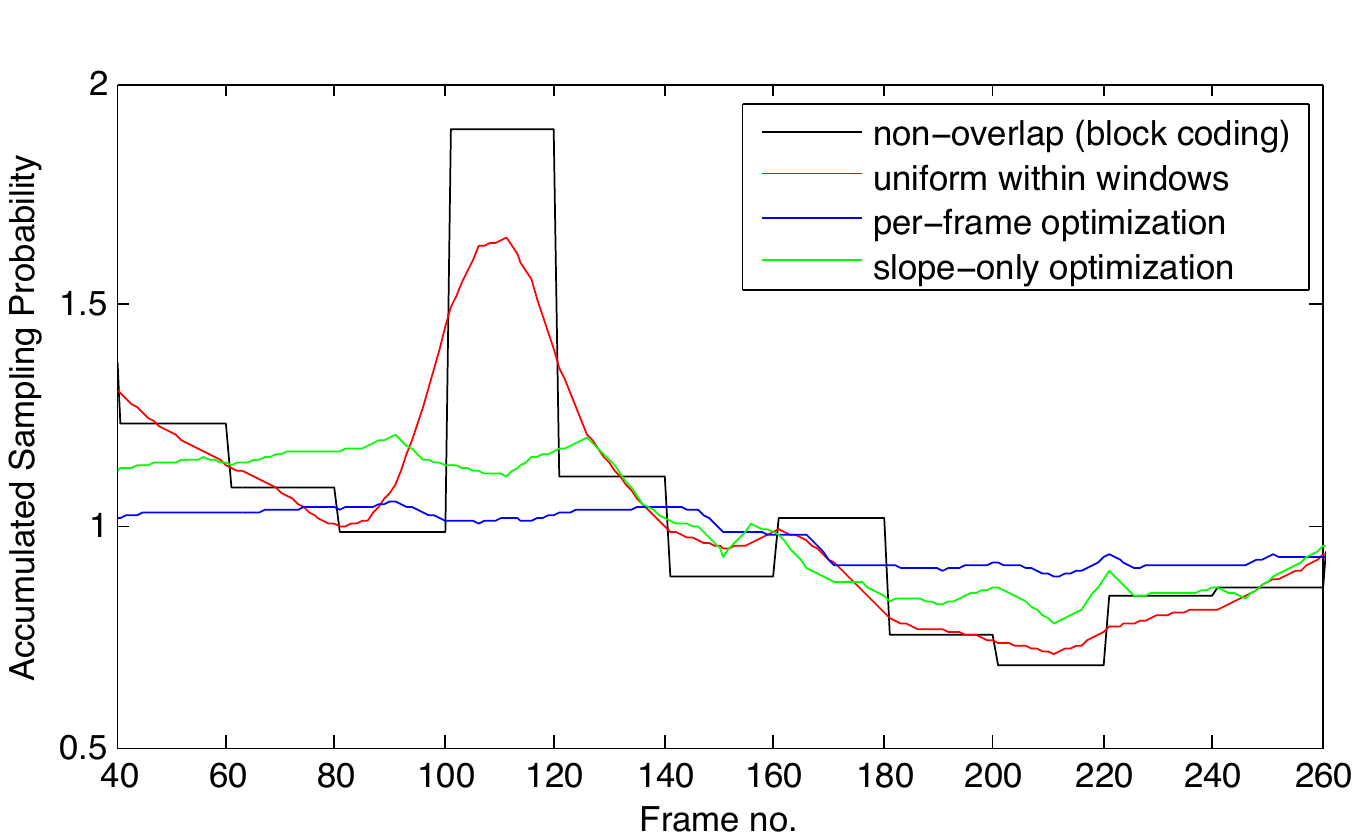}
		\caption{Comparison of the accumulated sampling probabilities using different sliding window schemes and optimization strategies for \textit{foreman}.}
		\label{fig:asps}
	\end{subfigure}
\caption{(a) The bit rate of \textit{foreman}, and (b) the results of ASP using different SWFC schemes. The figures show the frames in the range of 40 to 260. }
\label{fig:foremancompare}
\end{figure}

However, with the time-based sliding window, even if every window's sampling distribution is uniform, the overall sampling distribution may still be nonuniform. The reason is that the number of packets might be different for different frames. For example, as shown in Fig.~\ref{fig:bitrate}, the bit rate is not constant for the CIF video sequence \textit{foreman}. 
If the video is segmented into 20-frame blocks and the data rate of block coding fountain code is constant, however, as shown as the black line in Fig.~\ref{fig:asps}, the fluctuation of sampling probabilities is very huge. It is easy to understand that the probability is inversely proportional to the number of bits in that video block.

In SWFC scheme, instead of being related to only one window, the sampling probability of a frame is related to all the windows that covers it, as shown in (\ref{eq:ASP}).

\begin{equation}
P(t) = \sum\limits_{\omega \in \text{all windows cover frame }t} p_{\omega}^{pkt}(t),
\label{eq:ASP}
\end{equation}
where $p_{\omega}^{pkt}(t)$ denotes the average sampling probability of each packet in frame $t$ within the window $\omega$. So, $P(t)$ denotes the total probability of every packet in frame $t$ accumulated through all the sliding windows covering that frame, called \textit{accumulated sampling probability}, or ASP in the rest of this article. Here, because we assume that the bits within each frame have equal importance, it is assumed that all the packets in one frame have a same sampling probability, which leads to (\ref{eq:frmvspkt}).

\begin{equation}
p_{\omega}^{pkt}(t) = \frac{1}{s(t)}  p_{\omega}^{frm}(t)
\label{eq:frmvspkt},
\end{equation}
where $p_{\omega}^{frm}(t)$ denotes the total probability of the packets in frame $t$ to be sampled, within the window $\omega$. $s(t)$ is the number of packets in frame $t$ as defined in Table \ref{tab:video-define}.

For example, using a uniform-distribution sliding window (window size $W = 20$, step size $ \Delta t = 5$) to slide through the video sequence \textit{foreman}, we can obtain the ASP as shown as the red line in Fig.~\ref{fig:asps}. The ASPs shown here are normalized, so the average value of all the probabilities in one scheme is normalized to 1. Because the ASP forms non-uniform distribution, we know its coding efficiency is low for fountain code. 

Fortunately, the overlapping property of the sliding window provides a way to stabilize the ASP: the sampling probabilities within each window can be assigned unequally to achieve the overall uniformity of the ASP. 

Although selecting the best sampling distribution for each window is an optimization problem, we can still intuitively understand it as follows: if the vicinity of a window has relatively low bit rate and it will get higher in the future, in order to make the overall sampling distribution as homogeneous as possible, we do not want to ``waste'' the sampling opportunities on the imminent frames, which are already sampled in previous windows for too many times; instead, the encoder should sample more from the future side of the window, such that it could compensate the low sampling probability of the upcoming high bit rate frames.
 
In the rest of this section, we will introduce some solutions to this problem. To give a glimpse of what can the optimal window-wise sampling strategy do, the blue and green lines in Fig.~\ref{fig:asps} show the resulting ASPs using different optimization strategies. We can see that they are significantly more stable than the non-optimal schemes, which are in red and black.

\subsection{Per-frame Optimization Scheme}
\label{sec:per-frame}

In order to optimize the sampling distributions for all windows, we must know the video length $T$, window size $W$ and the number of packets in each frame $s(t)$ (or its vector form $\mathbf{s} = \begin{bmatrix}s(1) &  s(2) \cdots  s({T})\end{bmatrix}$). We define $p_{t}^{frm}(i)$ to denote the probability of sampling the packets in the $i^{\text{th}}$ frame of the window starting from the $t^{\text{th}}$ frame. As in (\ref{eq:frmvspkt}), the sampling probability for each packet in that frame within the window, $p_{t}^{pkt}(i)$, is defined as (\ref{eq:pktprob}).

\begin{equation}
p_{t}^{pkt}(i) = \frac{1}{s(t+i-1)}  p_{t}^{frm}(i)
\label{eq:pktprob}
\end{equation}

As in (\ref{eq:ASP}), the ASP for the $t^{\text{th}}$ frame is defined as (\ref{eq:probASP}). 

\begin{equation}
\label{eq:probASP}
\begin{split}
P(t) &= \sum\limits_{t_0 = t-W+1}^{t} p_{t_0}^{pkt}(t - t_0 + 1) \\
&= \frac{1}{s(t)}  \sum\limits_{t_0 = t-W+1}^{t} p_{t_0}^{frm}(t - t_0 + 1)
\end{split}
\end{equation}

For simplicity, this accumulation process does not consider the step sizes of $\Delta t$ other than $1$. Because both the video length $T$ and window size $W$ are defined to be integral multiples of $\Delta t$, if $\Delta t > 1$, all the parameters can be down-sampled by a factor of $\Delta t$. For example, the new $T' = \frac{1}{\Delta t}T$, $W' = \frac{1}{\Delta t}W$, $s'(t) = \sum\limits_{i=(t-1)\Delta t + 1}^{t\cdot \Delta t} s(i)$, and $\Delta t' = 1$, so (\ref{eq:pktprob}) and (\ref{eq:probASP}) still hold.

Let $x_{t,i} = p^{frm}_t(i)$ and make a matrix from them. We get the parameter matrix $\mathbf{A}$ as in (\ref{eq:Amatrix}).

\begin{equation}
\label{eq:Amatrix}
\mathbf{A} = 
 \begin{bmatrix}
 x_{1,1} & x_{1,2} & \cdots & x_{1,W} \\
 x_{2,1} & x_{2,2} & \cdots & x_{2,W} \\
 \vdots & \vdots & \ddots & \vdots \\
 x_{T-W+1,1} & x_{T-W+1,2} & \cdots & x_{T-W+1,W}
 \end{bmatrix}
\end{equation}

The number of rows is $W$ because each window has $W$ sampling probabilities. The number of columns is $T-W+1$ because there are $(T-W+1)$ windows in total (again, $\Delta t $ is assumed to be 1). Because every row in the matrix represents the probability distribution within a window, the elements in $\mathbf{A}$ must satisfy the constraints of (\ref{eq:constraints}).

\begin{equation}
\label{eq:constraints}
\begin{array}{{l}{l}}
{\sum\limits_{w = 1}^W {x_{t,w} = 1},} & { \forall t }\\
{ x_{t,w} \ge 0,} &{ \forall t,w}
\end{array}
\end{equation}

With this notation, (\ref{eq:probASP}) can be rewritten into a parameterized form as (\ref{eq:simpprobASP}).

\begin{equation}
P_{\mathbf{A}}(t) = \frac{1}{s(t)}  \sum\limits_{t_0 = t-W+1}^{t} x_{{t_0},t - t_0 + 1}
\label{eq:simpprobASP}
\end{equation}

The objective is to find the optimal parameter matrix $\mathbf{A}$, which minimizing the fluctuation of the ASPs, $P_{\mathbf{A}}(t)$. Because in this problem, the parameters to be optimized are the sampling probabilities for each frame of every window, we call this method the \textit{per-frame} optimization scheme.

\subsubsection{Problem Description}

Given the total number of frames $T$, the window size $W$, and number of packets in each frame $s(t)$, we want to find a set of parameters as in  (\ref{eq:Amatrix}), for which the mean square error of the sampling probabilities of all packets attains its minimum value. The optimization problem is defined in (\ref{eq:optperframe}).

\begin{equation}
\label{eq:optperframe}
\begin{array}{*{20}{l}{l}}
{\mathop {\arg\min }\limits_{\mathbf{A}}}&{\sum\limits_{t = W}^{T - W + 1} {{{\left( {{P_{\mathbf{A}}(t)} - \overline {P_{\mathbf{A}}} } \right)}^2}}} & {}\\
{s.t.}&{\sum\limits_{w = 1}^W {x_{t,w} = 1},} & { \forall t }\\
{}&{ x_{t,w} \ge 0,} &{ \forall t,w}
\end{array}
\end{equation}

where ${\overline {{P_{\mathbf{A}}}}  = \frac{1}{{T - 2W + 2}}\sum\limits_{t = W}^{T - W + 1} {{P_{\mathbf{A}}(t)}} }$.

It should be noted that the range of frames we want to stabilize is from $W$ to $W-T+1$. Because the frames in that range are all covered by exactly $W$ sliding windows, they are deemed as stable frames. On the other hand, the frames before $W$ or after $W-T+1$ are covered by less than $W$ sliding windows, so they are considered to be warm-up/cool-down frames, and not be counted as the targets of the optimization.

\subsubsection{Solution}

If the conditions of ${ x_{t,w} \ge 0}$ are ignored, this optimization problem can be solved using Lagrange multiplier. Otherwise, it can be solved by Karush-Kuhn-Tucker (KKT) conditions.

\begin{figure}[]
\centering
    \includegraphics[width= 0.35\textwidth]{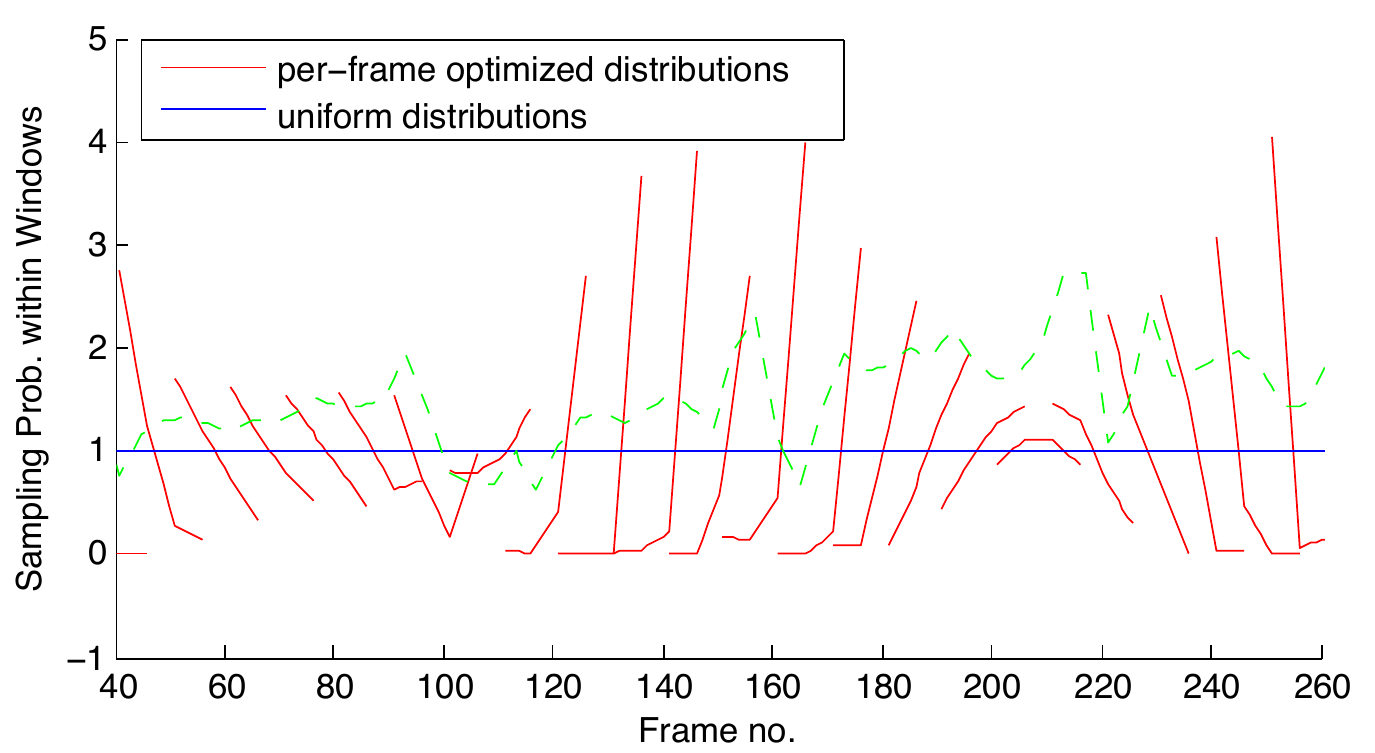}
    \caption{The optimization result of sampling distributions for each window of \textit{foreman} using per-frame optimization scheme. }
    \label{fig:foreman-per-frame-opt}
\end{figure}

An example of the optimization result is shown in Fig.~\ref{fig:foreman-per-frame-opt}. It is the optimization result of sampling distributions for each window of CIF sequence \textit{foreman} using per-frame optimization scheme. Window size $W=20$ and step size $\Delta t=5$. Because there are too many windows to be clearly shown in one figure, only a fraction of the windows is presented here. The probabilities are normalized. The trend of the bit rate, which is represented by dashed green line, is also plotted in the figure, in order to indicate the relationship between bit rate and optimization results. The blue line in Fig.~\ref{fig:asps} shows the resulting ASP using this per-frame optimization strategy.

\subsubsection{Computational Complexity}

Because there are $\frac{W \times (T-W+\Delta t)}{\Delta t^2}$ variables to optimize and $\frac{T-W}{\Delta t} + 1$ conditions for Lagrange multiplier (if using KKT conditions, there are $\frac{(W+\Delta t) \times (T-W+\Delta t)}{\Delta t^2} $ conditions), the optimization process yields the system of equations with $\frac{(W+\Delta t) \times (T-W+\Delta t)}{\Delta t^2}$ equations (or $\frac{(2 W+\Delta t) \times (T-W+\Delta t)}{\Delta t^2} $ equations in KKT conditions). Assuming that $T \gg W \gg \Delta t$, if we omit constant factors and lower order terms, the solution of both KKT conditions and Lagrange multiplier involves the generation of a parameter matrix of $\frac{T \cdot W}{\Delta t^2} \times \frac{T \cdot W}{\Delta t^2}$ and the computation of its inverse matrix. As a result, the computational complexity is  $O\Big((\frac{T \cdot W}{\Delta t^2})^3\Big)$.

\subsection{Slope-only optimization scheme}
\label{sec:slope-only}

Although optimizing the sampling distribution for each frame within every window yields the most optimal solution in terms of minimizing the fluctuation of sampling probabilities between frames, per-frame optimization is unrealistic in practical designs. First of all, there are too many parameters to be optimized. The computational complexity is  $O\Big((\frac{T \cdot W}{\Delta t^2})^3\Big)$, which is too high for large $T$ or $W$. Secondly, in order to reconstruct the coded packets on the decoder side, the encoder must tell the receiver what sampling distribution is used in each window, by explicitly including every frame's sampling probability in the packet header. That will introduce a large overhead in the packet header. Since bigger packets are more vulnerable to channel noise, including too much information in headers will increase packet loss rate in wireless networks. As a result, a more concise description for the sampling distributions is needed for the practical designs, so they can be obtained with lower computational complexity, and be transmitted in the headers with less bits.

We introduce a \textit{slope-only} description for the sampling distributions. It requires only one parameter -- \textit{slope factor}, denoted as $a$, to control the shape of the distribution. However, it should be noted that using less bits will inevitably lose the precision of describing the sampling distributions. Therefore, compared to the optimal performance that can achieved by using per-frame description, slope-only description may result in suboptimal performance.

\begin{figure}[]
	\centering
	\includegraphics[width= 0.45\textwidth]{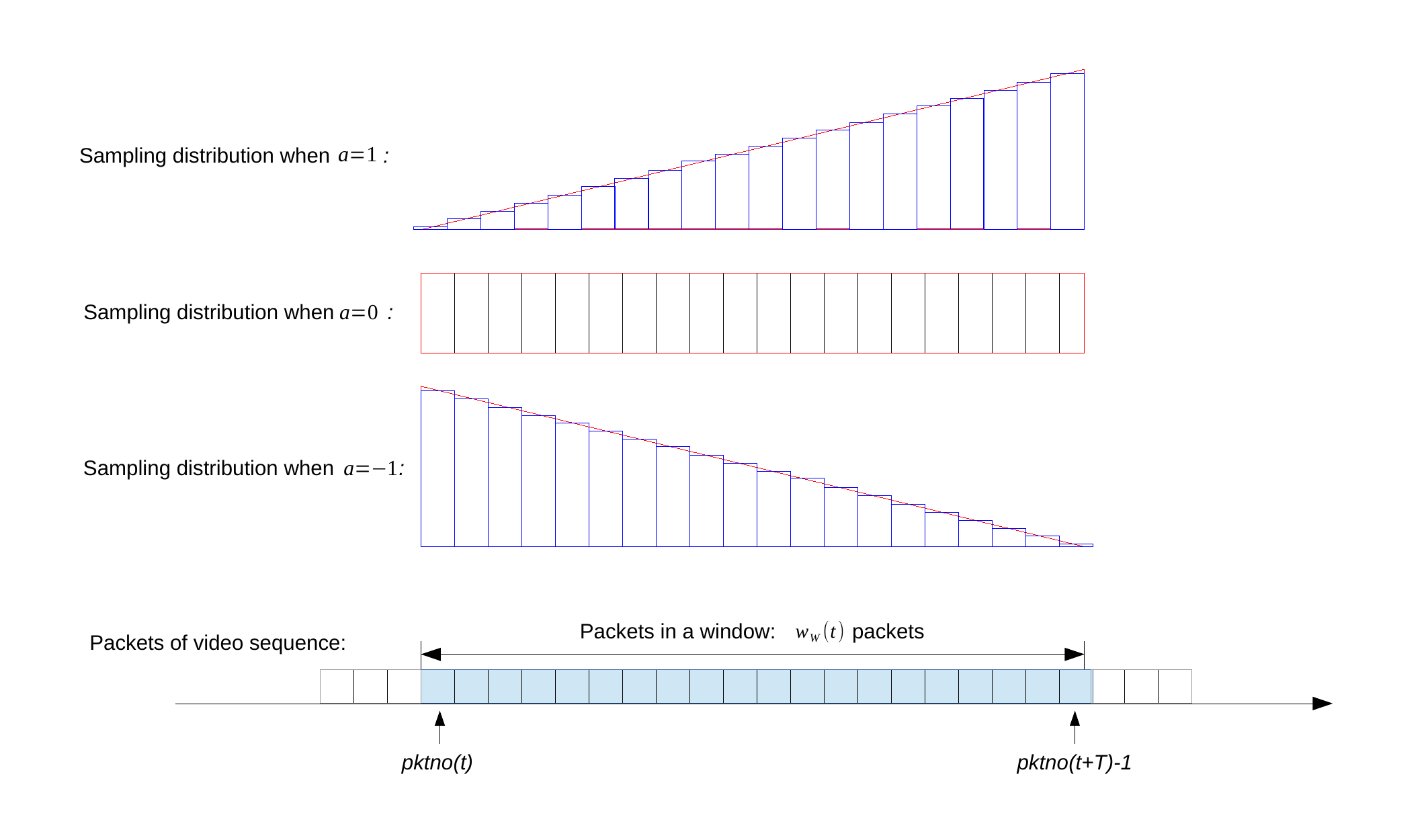}
	\caption{The sampling distributions when slope factor $a=1$, $a=0$, and $a=-1$.}
	\label{fig:slope}
\end{figure}

The slope factor is a real number, and it ranges from $-1$ to $1$. The distribution functions are defined to be linear functions, and the slope factor only controls the slopes of them. 
We do not want any of the packet's probability to be 0, because in that case, the effective window size will shrink. As a result, we define that when the slope factor $a = 1$, the distribution function of the packets starts from 0 and increases linearly, forming a \textit{forward triangular distribution}, as shown in the top of Fig.~\ref{fig:slope}; when the slope factor $a=0$, the distribution function is a \textit{uniform distribution}, as shown in the second line of Fig.~\ref{fig:slope}; when the slope factor $a=-1$, the distribution function is the reverse of that in $a=1$, or a \textit{backward triangular distribution}, as shown in the third line of Fig.~\ref{fig:slope}. Therefore, the distribution functions of all the slope factor values in the middle are continuously defined.

\begin{figure}[]
	\centering
	\includegraphics[width= 0.35\textwidth]{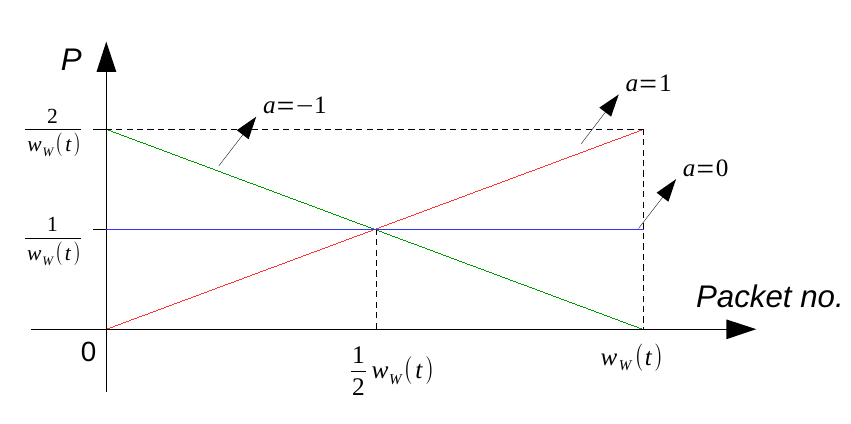}
	\caption{The distribution functions when slope factor $a=1$, $a=0$, and $a=-1$.}
	\label{fig:function}
\end{figure}

As defined in table \ref{tab:fountain-define}, for time $t$ in video sequence, the number of packets in the window is $w_{W}(t)$. For each window, a linear distribution function can be defined over the interval $[0,w_{W}(t)]$. Because the integration of the function in $[0,w_{W}(t)]$ must be 1, we can get the definitions of the lines for different slope factors $a$. When slope factor $a=1$, it passes points $(0,0)$ and $\Big(w_{W}(t), \frac{2}{w_{W}(t)}\Big)$, as the red line shown in Fig.~\ref{fig:function}; when slope factor $a=-1$, it passes points $\Big(0, \frac{2}{w_{W}(t)}\Big)$ and $\Big(w_{W}(t), 0\Big)$, as the green line shown in Fig.~\ref{fig:function}. The lines for all slope factors will always pass the point $\Big(\frac{1}{2} w_{W}(t), \frac{1}{w_{W}(t)}\Big)$. As a result, the distribution function given any $a$ and $t$ is (\ref{eq:slopefunction}).

\begin{equation}
	\label{eq:slopefunction}
	y = \frac{2a}{w_{W}^2(t)}x + \frac{1-a}{w_{W}(t)} , \qquad x \in [0,w_{W}(t)]
\end{equation}

\begin{figure}[]
	\centering
	\includegraphics[width= 0.35\textwidth]{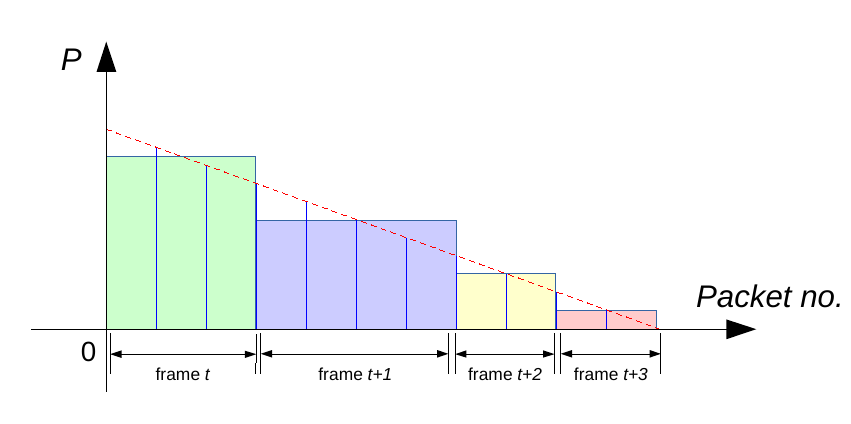}
	\caption{An example of sampling distribution for each frame within a window. The window contains four frames, and the sampling probabilities of the packets within a same frame should be the same.}
	\label{fig:framefunction}
\end{figure}

As stated in section \ref{sec:per-frame}, the sampling probabilities of the packets within a same frame should be the same. As a result, the probabilities of sampling one frame should be grouped together, and the actual sampling probability of each packet should be the average value of all packets in its frame. As the example shown in Fig.~\ref{fig:framefunction}, there are four frames, each of which contains 3, 4, 2 and 2 packets respectively. The actual sampling probability for each packet is the average value of the packets in the interval of its frame. Given slope factor $a$, the probability of sampling the $i^{\text{th}}$ frame in the window starting from the $t^{\text{th}}$ frame, denoted as $p^{frm}_{t,a}(i)$, is then defined as (\ref{eq:framefunction}).

\begin{equation}
	\label{eq:framefunction}
	\begin{split}
		p^{frm}_{t,a}(i) & = \int_{pkt(t,i-1)}^{pkt(t,i)} \Big( \frac{2a}{w_{W}^2(t)}x + \frac{1-a}{w_{W}(t)} \Big) dx \\
		& =  \Bigg(\frac{2a}{w_{W}^2(t)}\Big(pkt(t,i)-\frac{s(t+i-1)}{2}\Big) + \frac{1-a}{w_{W}(t)} \Bigg) \\ 
		& \qquad \times s(t+i-1), \\
		i& = 1, 2, ... , W
	\end{split}
\end{equation}

where $pkt(t,i)$ is defined in table \ref{tab:video-define}. The second equality holds because the distribution function is a linear function, and the average value is taken at the middle point of each interval.

As in (\ref{eq:frmvspkt}), given slope factor $a$, the probability of each packet to be sampled in the $i^{\text{th}}$ frame within the $t^{\text{th}}$ window, denoted as $p^{pkt}_{t,a}(i)$, is the average value of the distribution function in the interval $[pkt(t,i-1),pkt(t,i)]$, which is defined in (\ref{eq:samplefunction}).

\begin{equation}
	\label{eq:samplefunction}
	\begin{split}
		p^{pkt}_{t,a}(i) &= \frac{p^{frm}_{t,a}(i)}{s(t+i-1)}\\
		&= \frac{2a}{w_{W}^2(t)}\Big(pkt(t,i)-\frac{s(t+i-1)}{2}\Big) + \frac{1-a}{w_{W}(t)} ,\\
		\text{for }i&= 1, 2, ... , W.
	\end{split}
\end{equation}

%
%
%
%
%

As in (\ref{eq:ASP}), the ASP for each packet in frame $t$, denoted as ${P_{\mathbf{a}}(t)}$ is defined in (\ref{eq:accumslope}).

\begin{equation}
	\label{eq:accumslope}
	{{P_{\mathbf{a}}(t)} = \sum\limits_{{t_0} = t - W + 1}^t {p^{pkt}_{{t_0},{a_{t_0}}}(t - {t_0} + 1) } },
\end{equation}
\begin{equation}
	\label{eq:vector}
	\mathbf{a} = 
	\begin{bmatrix}
		a_1 &  a_2 \cdots  a_{T-W+1}
	\end{bmatrix},
\end{equation}
where $\mathbf{a}$ denotes the set of slope factors for all windows in the video sequence (from frame $1$ to frame $(T-W+1)$ ) as in  (\ref{eq:vector}). Again, for simplicity, the accumulation process does not consider the step sizes of $\Delta t$ other than $1$.

We can rewrite (\ref{eq:accumslope}) for clearer notations as in (\ref{eq:simpleaccumslope}).

\begin{equation}
	\label{eq:simpleaccumslope}
	{{P_{\mathbf{a}}(t)} = \mathbf{d_1} (t) \cdot \mathbf{a} + d_2(t) },
\end{equation}
where ``$ \cdot $'' denotes the dot product of the two vectors of $(T-W+1)$ elements, and

\begin{equation}
	\label{eq:wheresimpleaccumslope}
	\begin{split}
		\mathbf{d_1}(t) & = 
		\begin{bmatrix}
			d_{1}(t,1) &  d_{1}(t,2) \cdots  d_{1}(t,T-W+1)
		\end{bmatrix};\\
		d_1(t,i) & = 
		\begin{cases}
			\frac{2 \cdot pkt(i,t-i+1)-{s(t)}}{w_{W}^2(i)} - \frac{1}{w_{W}(i)}, \hfill \mbox{if $i \in [t-W+1, t]$;} \\
			0, \hfill \mbox{otherwise;}\\
		\end{cases}\\
		d_2(t)   & =  \sum\limits_{t_0=t-W+1}^{t} \frac{1}{w_{W}(t_0)}.
	\end{split}
\end{equation}

From (\ref{eq:wheresimpleaccumslope}) we can see that $\mathbf{d_1}$ and $d_2$ are only relevant to $\mathbf{s}$, $W$ and $t$, but not influenced by $\mathbf{a}$. 

With the equations defined above, we can describe the optimization problem as follows.

\subsubsection{Problem Description}

Given the total number of frames $T$, the window size $W$, and number of packets in each frame $s(t)$, we want to find a set of slope factors as in  (\ref{eq:vector}), for which the mean square error of the sampling probabilities of all packets attains its minimum value. The optimization problem is defined in (\ref{eq:optslope}).

\begin{equation}
	\label{eq:optslope}
	\begin{array}{*{20}{c}}
		{\mathop {\arg\min }\limits_{\mathbf{a}} }&{\sum\limits_{t = W}^{T - W + 1} {{{\left( {{P_{\mathbf{a}}(t)} - \overline {P_{\mathbf{a}}} } \right)}^2}} }\\
		{s.t.}&{-1 \le a_t \le 1, \qquad \forall t}
	\end{array}
\end{equation}

where ${\overline {P_{\mathbf{a}}}  = \frac{1}{{T - 2W + 2}}\sum\limits_{t = W}^{T - W + 1} {P_{\mathbf{a}}(t)} }$. The range of frames we want to stabilize is also from $W$ to $W-T+1$, for the same reason as stated in the per-frame condition.

\subsubsection{Solution}

\begin{figure}[]
	\centering
	\includegraphics[width= 0.35\textwidth]{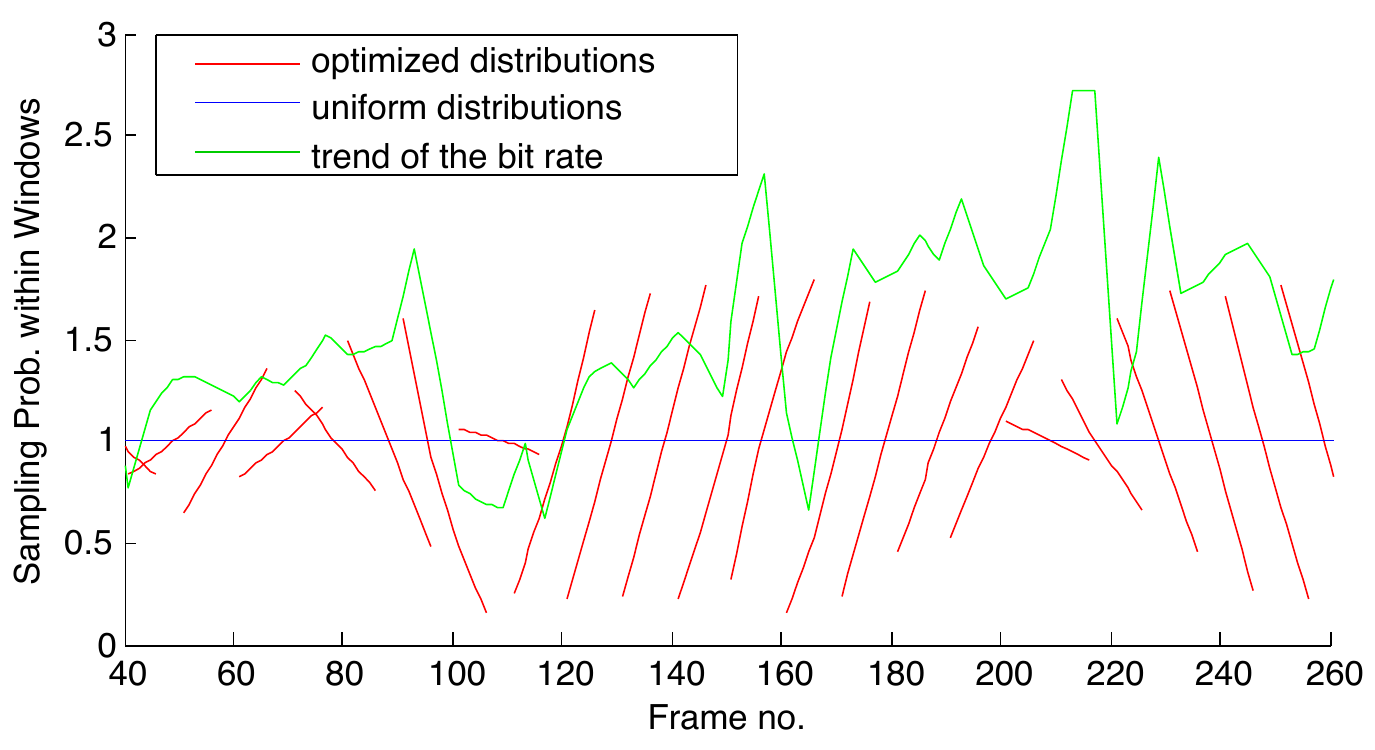}
	\caption{The optimization result of sampling distributions for each window of \textit{foreman} using slope-only optimization scheme.}
	\label{fig:foreman-slope-only-opt}
\end{figure}

As in the per-frame scheme, it can be solved by KKT conditions. The result of each window's sampling distribution using slope-only optimization is shown in Fig.~\ref{fig:foreman-slope-only-opt} with the same settings as in Fig.~\ref{fig:foreman-per-frame-opt}. The green line in Fig.~\ref{fig:asps} shows the resulting ASP using this slope-only optimization strategy. We can observe that, in terms of stability of ASP, slope-only scheme yields worse result than per-frame scheme.

\subsubsection{Computational Complexity}

Because there are $\frac{T-W}{\Delta t} + 1$ variables to optimize and $2\times (\frac{T-W}{\Delta t} + 1)$ conditions for KKT conditions, the optimization process yields the system of equations with $3\times (\frac{T-W}{\Delta t} + 1)$ equations. Assuming that $T \gg W \gg \Delta t$, if we omit constant factors and lower order terms, the solution involves the generation of a parameter matrix of $\frac{T}{\Delta t} \times \frac{T }{\Delta t}$ and the computation of its inverse matrix. As a result, the computational complexity is  $O\Big((\frac{T}{\Delta t})^3\Big)$. Compared to that of per-frame scheme, the computational complexity of slope-only scheme is lowered by the factor of $(W/\Delta t)^3$.

\section{System Design}
\label{sec:system}

In this section, we design a practical system for the delay-aware fountain code scheme. From the acronyms of the scheme, we name our protocol as DAF. 

Because of the reasons we stated in section \ref{sec:slope-only}, we design DAF protocol based on the slope-only description and optimization scheme. 

\subsection{Warm-up/Cool-down Period}
\label{sec:warmup}

\begin{figure}[]
	\centering
	\includegraphics[width= 0.35\textwidth]{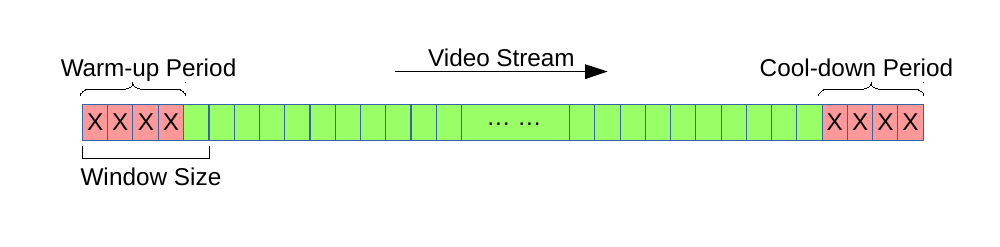}
	\caption{The definition of warm-up/cool-down periods.}
	\label{fig:warm-up}
\end{figure}

If the window always contains $W$ frames, and it slides from the $1^{\text{st}}$ frame to the $T^{\text{th}}$ frame at the speed of $\Delta t = 1$ , it is easy to realize that all the frames will be covered in $W$ windows, except for the first $W-1$ frames and the last $W-1$ frames. Namely, the first and the last $i^{\text{th}}$ frame will be covered in $i$ windows when $i \le W - 1$. We call those two periods as \textit{warm-up} and \textit{cool-down} periods (W/CP), as illustrated in Fig.~\ref{fig:warm-up}, since they are undersampled and yield unstable decoding ratio.

In our implementation, before the actual SWFC begins, both encoder and decoder will obtain the length of W/CP. The encoder will fill these two periods with padding characters, and the decoder will do the same and automatically mark those packets as decoded. Then, the SWFC is performed. The detailed procedures of encoder and decoder will be introduced in section \ref{sec:encode} and \ref{sec:decode}. Also, for the sake of fairness, the pseudo-decoded padding packets in W/CP should not be counted as being decoded in evaluation, since they do not contain any useful information.

\subsection{Packet Structure}
\label{sec:packet-structure}

\begin{figure}[ht]
	\centering
	\includegraphics[width=0.24\textwidth]{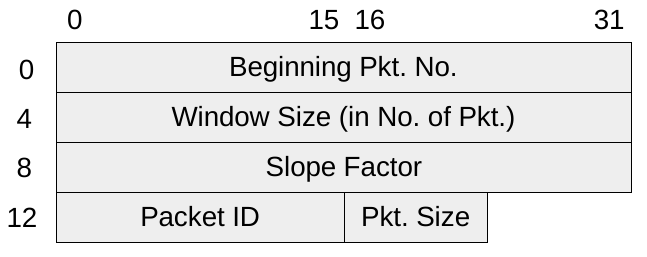}
	\caption{The header structure of a DAF packet.}
	\label{fig:headerstr}
\end{figure}

The structure of a DAF packet header is shown in Fig.~\ref{fig:headerstr}. The payload of a DAF packet is coded, and its length is given in the header. The total size of header is 15 bytes. It includes the starting packet position of the window ($StartP$), the size of current window in the unit of number of packets ($WSize$), the slope factor used in current window ($SlopeF$), packet ID ($PacketID$) and packet size ($P$). 

The data length of $SlopeF$ determines the precision of the slope factors used in generating sampling distribution. In our protocol design, we use 4 bytes as the length of it, which stores a real number as the float type in C++. $PacketID$ starts from 1, and will be increased by 1 every time a coded packet is sent. It serves the similar purpose as in fountain code, which is the random seed for generating degrees and sampling packets. 

If a user does not need the sampling distribution optimization due to limited computational power, $SlopeF$ field can be set to 0, which means uniform distribution, to have the low complexity version of DAF (DAF-L).

\subsection{DAF Encoder}
\label{sec:encode}

\begin{figure}[]
\centering
    \includegraphics[width= 0.5\textwidth]{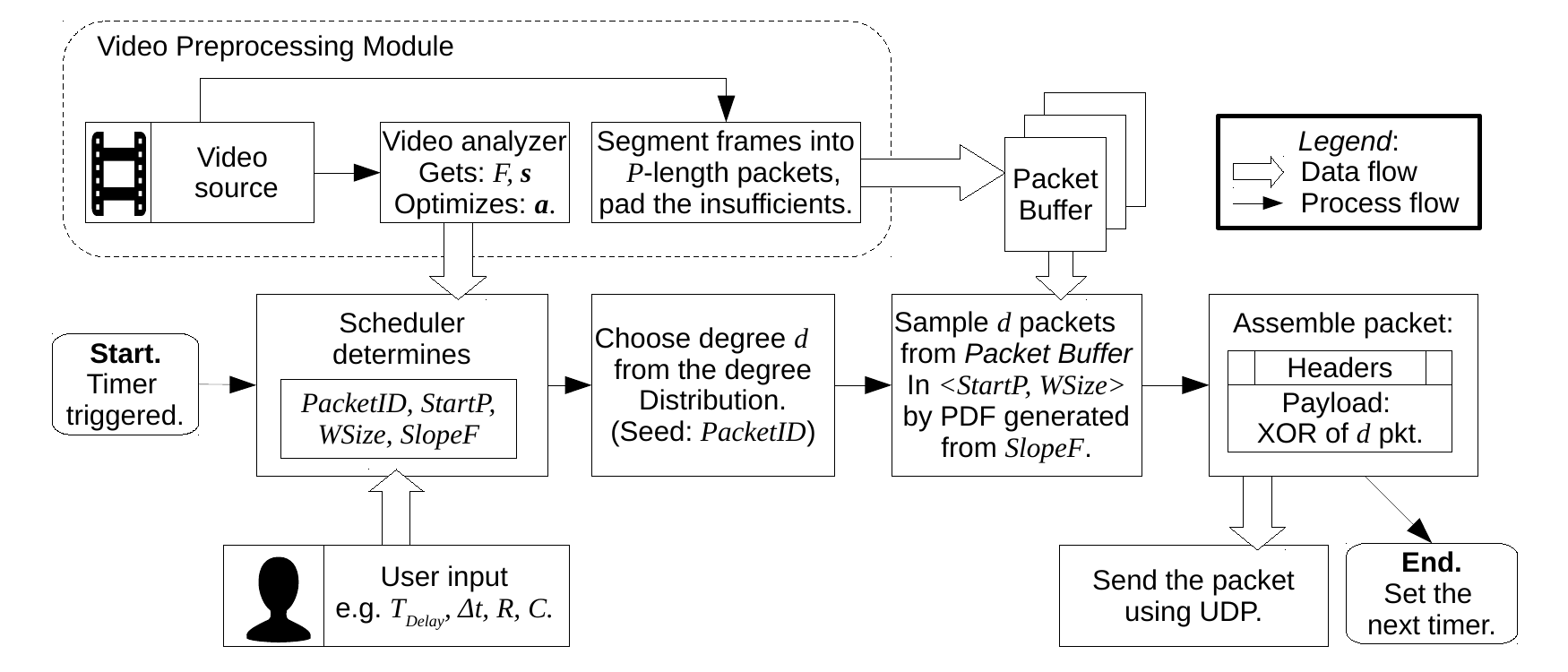}
    \caption{The flowchart of the DAF encoder.}
    \label{fig:encoder-flowchart}
\end{figure}

The system design of DAF encoder is shown as a flowchart in Fig.~\ref{fig:encoder-flowchart}. Beforehand, the coding parameters, degree distributions and W/CP are already obtained by the encoder. The system takes two sets of input: the parameters assigned by user (e.g. $T_{Delay}$,$\Delta t$,$R$,$C$ ) and the video source.

The video source feeds the system with streamed video data, and it is first processed by the \textit{video preprocessing module}, as shown in the dotted box. This module gets the information such as $F$, $\mathbf{s}$, $N_{GOP}$, and optimizes the slope factors $\textbf{a}$. It also segments the data from each frame (or GOP) in to several $P$-byte packets, and pads the insufficient packets to $P$ bytes. It puts the segmented video packets in the packet buffer.

The middle row of the flowchart describes the encoding algorithm of DAF system. After the procedure is triggered by the timer, the scheduler will determine whether to move the window to the next position, according to the parameters and the current status. If not, $StartP$, $WSize$ and $SlopeF$ remain the same as last sent packet; if the window slides, let $StartP= StartP + \Delta t$, $WSize = w_{W}(StartP)$, and $SlopeF = a(StartP)$. In both cases, let $PacketID = PacketID+1$.

In the next step, a degree $d$ is chosen according to the degree distribution, like that in LT codes. Then, $d$ packets are sampled from the packet buffer in the range confined by $StartP$ and $WSize$. Each window's packet-wise sampling distribution is generated by (\ref{eq:samplefunction}), given $SlopeF$. The bit-wise XOR of these $d$ original packets is obtained as the payload of current coded packet.

At last, the parameters and the payload are assembled as an APP layer packet, according to the structure shown in Fig.~\ref{fig:headerstr}. The packet will be sent using UDP. Last but not least, the program will set the timer to trigger the procedure again according to the frequency of sending packets, which is determined by parameters such as $F$,$P$,$R$,$C$, etc.

\subsection{DAF Decoder}
\label{sec:decode}

\begin{figure}[]
\centering
    \includegraphics[width= 0.5\textwidth]{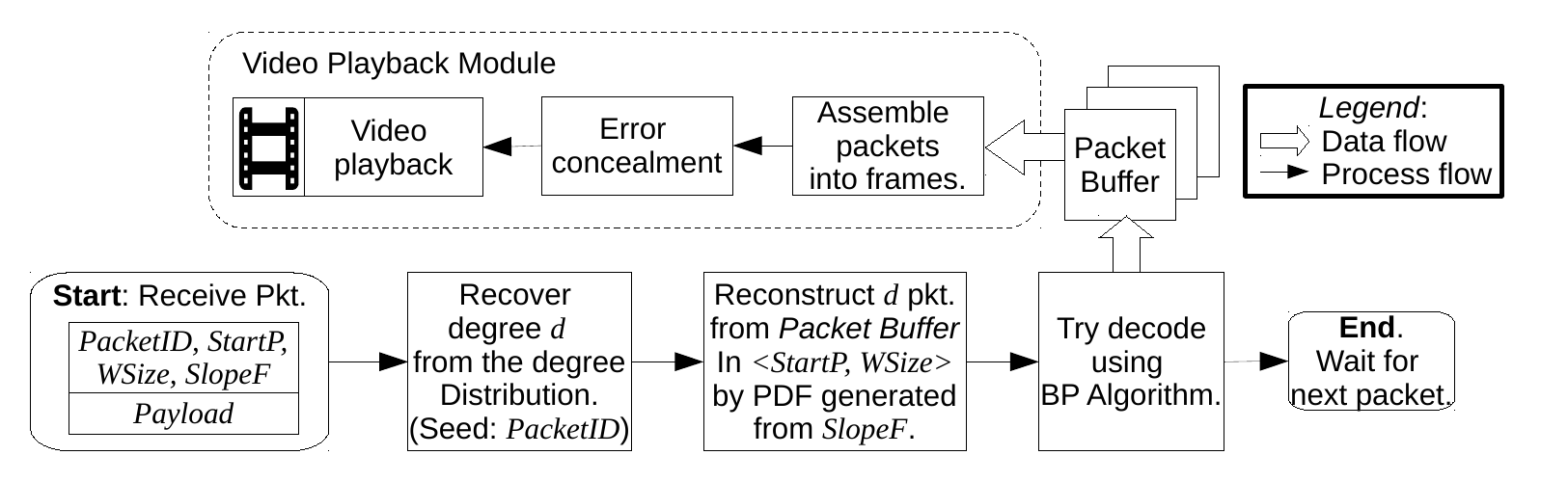}
    \caption{The flowchart of the DAF decoder.}
    \label{fig:decoder-flowchart}
\end{figure}

The system design of DAF decoder is shown as a flowchart in Fig.~\ref{fig:decoder-flowchart}. Also, the coding parameters, degree distributions and W/CP are already obtained by the decoder. The procedure starts when a coded packet is received.

The decoding procedure is basically the reverse of encoding procedure. Having $StartP$, $WSize$, $SlopeF$ and $PacketID$, the degree $d$, the sampling distribution and the composition of the coded packet can be reconstructed. They are fed into a belief propagation (BP) decoder, which tries to decode the original packets. The decoded packets are stored in the packet buffer. 

The \textit{video playback module} requests packets from the packet buffer as the time goes. First, the packets are re-assembled into frames (or GOPs). If a packet has not been decoded yet when it is requested, it is considered as a packet loss. If this happens, image processing techniques such as error concealment may be performed to fix it before playing it. 

\subsection{General Framework of Fountain Code Systems}
\label{sec:general}

It is worth mentioning that the framework of DAF system is a generalization of many existing schemes based on fountain codes. Different fountain code schemes can be easily implemented by changing settings and modules in DAF system, but the protocol does not need to be changed.

For example, if a user does not need the sampling distribution optimization due to limited computational power, $SlopeF$ can be set to 0, which means uniform distribution, to have a low complexity version of DAF (DAF-L); the original fountain code can be viewed as a special case when $W=T$, and let the timer continually send the coded packets until an ACK is received; the block coding schemes can also be viewed as special case when $\Delta t = W$; furthermore, the sliding window schemes with packet-based window size, like \cite{FirstSliding}, is a special case with fixed $WSize$; finally, expanding window \cite{ExpandWindow,ExpandWindow2} can be viewed as another special case if we modify the scheduler of the encoder, by fixing $StartP$.

As a result, the proposed system enjoys the flexibility to meet different requirements.

\section{Simulation Experiments and Performance Evaluation}
\label{sec:experiments}


\subsection{Simulator Setup}
\label{sec:simulator}

We conduct the simulation experiments on Common Open Research Emulator (CORE) \cite{CORE} and Extendable
Mobile Ad-hoc Network Emulator (EMANE) \cite{EMANE}. The former provides virtualization on application (APP), transport (UDP or TCP) and network (IP) layer controlled by a graphical user interface, and the latter provide high-fidelity simulation for link (MAC) layer and physical (PHY) layer. The working environment is set up on Oracle$^\text{\circledR}$ VM VirtualBox virtual machines.

We use CORE to emulate the topology of the virtual network and the relay nodes. Two VMs are connected to the virtual network as a source (or encoder/sender) node running the client application, and a destination (or decoder/receiver) node running the server application. A video is streamed from client to server using different schemes. 

EMANE is used for emulation of IEEE 802.11b on PHY and MAC layer of each wireless node. Because of the forward error correction (FEC) nature of fountain code, we disable the retransmission mechanism of 802.11b for all fountain-code-based schemes. For the simplicity of performance evaluation, we also disable the adaptive rate selection mechanism of 802.11b, and only allow the 11 Mbps data rate to be used. Ad-hoc On-Demand Distance Vector (AODV) protocol is used for routing.


\subsection{Performance Metric}
\label{sec:eval-crit}

In our work, we use packet decoding ratio to evaluate the performance of the schemes, since higher packet decoding ratio implies higher visual quality of video. It is worth noting that the evaluation criteria of delay-aware multimedia streaming is different from the file transfer applications, and it is commonly overlooked by existing SWFC schemes. In delay-aware applications, if a packet is decoded after its playback time, it has to be counted as a packet loss for the video decoder, since the player does not rewind the video. 

As a result, we introduce the metric of \textit{in-time decoding ratio} ($IDR$), which only counts a decoded packet as ``in-time'' decoded when it is within the current window. Comparatively, \textit{file decoding ratio} ($FDR$) means the percentage of total decoded packets after the complete coding session finishes. For SWFC schemes, there is always $FDR \ge IDR$; for block coding, $FDR = IDR$.

\subsection{Performance Evaluation}
\label{sec:perfeval}

We conduct experiments for the following cases: (i) one hop with no node mobility (fixed topology) under various delay requirements and code rates, (ii) various number of hops with no node mobility (fixed topology) under fixed packet loss rate per hop, (iii) two hops with a moving relay node (dynamic topology). We implement six schemes for comparisons, which are abbreviated as follows:

\begin{enumerate}
\item \textit{DAF}: the proposed delay-aware fountain code protocol as introduced in section \ref{sec:system}.

\item \textit{DAF-L}: it is the low complexity version of DAF scheme. It is DAF without using the optimized window-wise sampling distribution, as proposed in section \ref{sec:packet-structure}. 

\item \textit{S-LT}: the sliding window LT code from \cite{FirstSliding}. 

\item \textit{Block}: the block coding for fountain codes. 

\item \textit{Expand}: this is the expanding window scheme of \cite{ExpandWindow}. 


\item \textit{TCP}: this scheme uses TCP protocol to stream video. In order to add delay awareness, the video file is also segmented into the blocks like in ``Block'' scheme, but they are sent using TCP. For the sake of fairness, the maximum data rate is limited to the same amount as required by the SWFC schemes. 

\end{enumerate}

All the five fountain-code-based schemes use the following parameter setting: the packet size $P=1024$ bytes; for degree distribution, let $\delta = 0.02$, $c=0.4$ (as defined in \cite[Definition 11]{LTCodes}), so LT code can get good average performances.

Several benchmark CIF test sequences, provided by \cite{sequence}, are used for our evaluation. They are coded into H.264/AVC format using \textit{x264} \cite{x264}, encapsulated into ISO MP4 files using \textit{MP4Box} \cite{mp4box}, and streamified by \textit{mp4trace} tool from \textit{EvalVid} tool-set \cite{evalvid}. The coding structure is IPPP, which contains only one I-frame (the first frame) and no B-frame, and all the rest are P-frames. Let $N_{GOP} = 1$. For the sake of clarity, all the delays shown in the experiments are in the unit of seconds. Because the frame rate for all sequences are 30 frames per second, it is easy to convert the unit between seconds and number of frames. Denote by $C$ and $T_{Delay}$ the code rate and the tolerable delay, respectively.

We conduct 20 experiments for each setting with different random seeds, and take the median value of them as the performance measure. Two results are shown for each set of experiment: in-time decoding ratio ($IDR$) and file decoding ratio ($FDR$).

\subsubsection{Case 1: One hop with no node mobility} 
\label{sec:case1}

In this case, there are two nodes in the network: a source node and a destination node. The communication path from the source to the destination has one hop. The distance between the two nodes is carefully set so that the packets with 1024-byte payload will have $10\%$ packet loss rate ($PLR$). Let $\Delta t=1$.

\begin{figure}[]
	\centering
	\includegraphics[width= 0.35\textwidth]{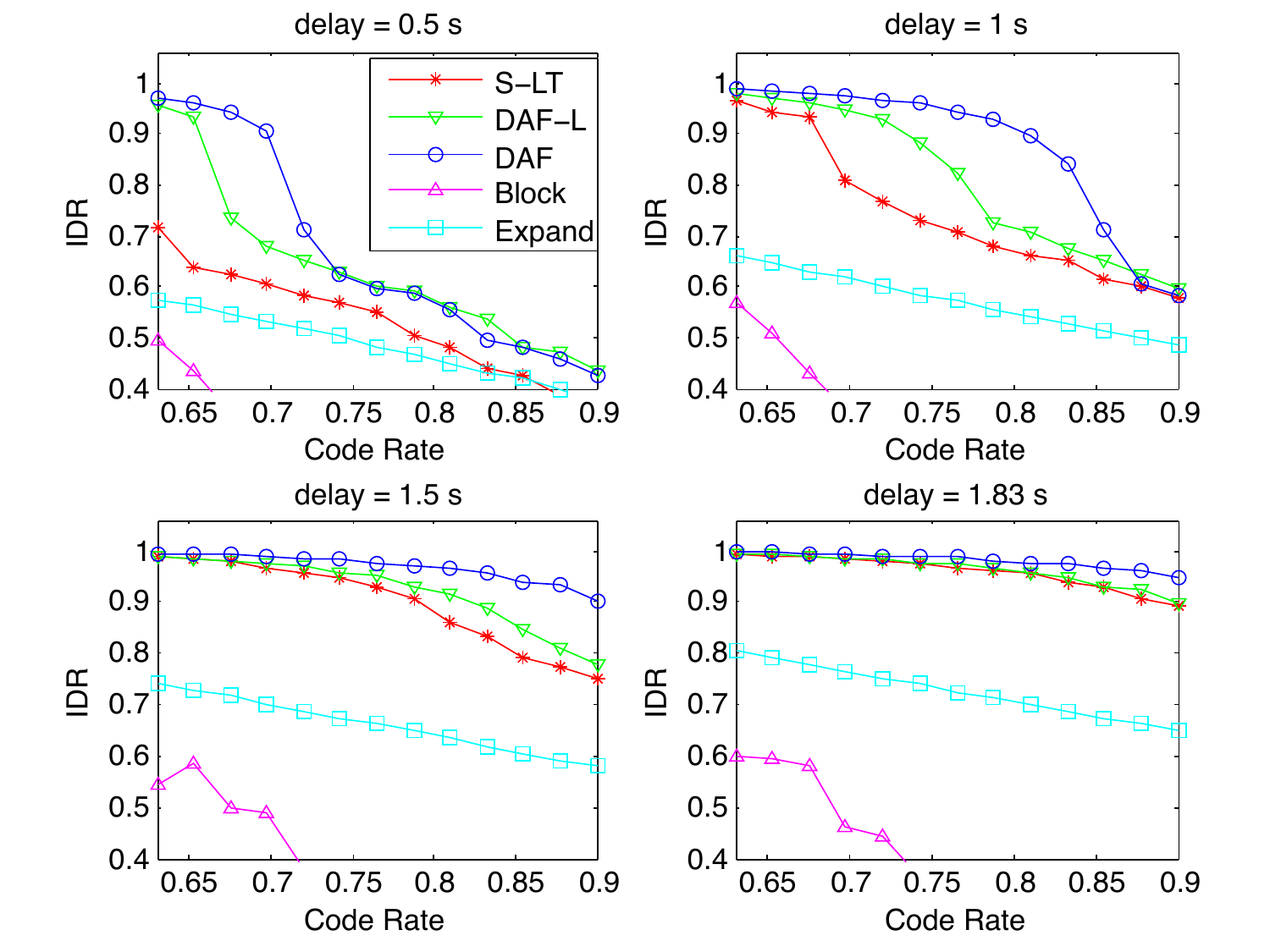}
	\caption{Relations of IDR vs. code rate of CIF sequence \textit{foreman} when $T_{Delay}=$ 0.5, 1, 1.5, and 1.83 seconds. Five sliding window schemes are compared. $PLR=10\%$. $\Delta t=1$. }
	\label{fig:resultDelay}
\end{figure}
\begin{figure}[]
	\centering
	\includegraphics[width= 0.35\textwidth]{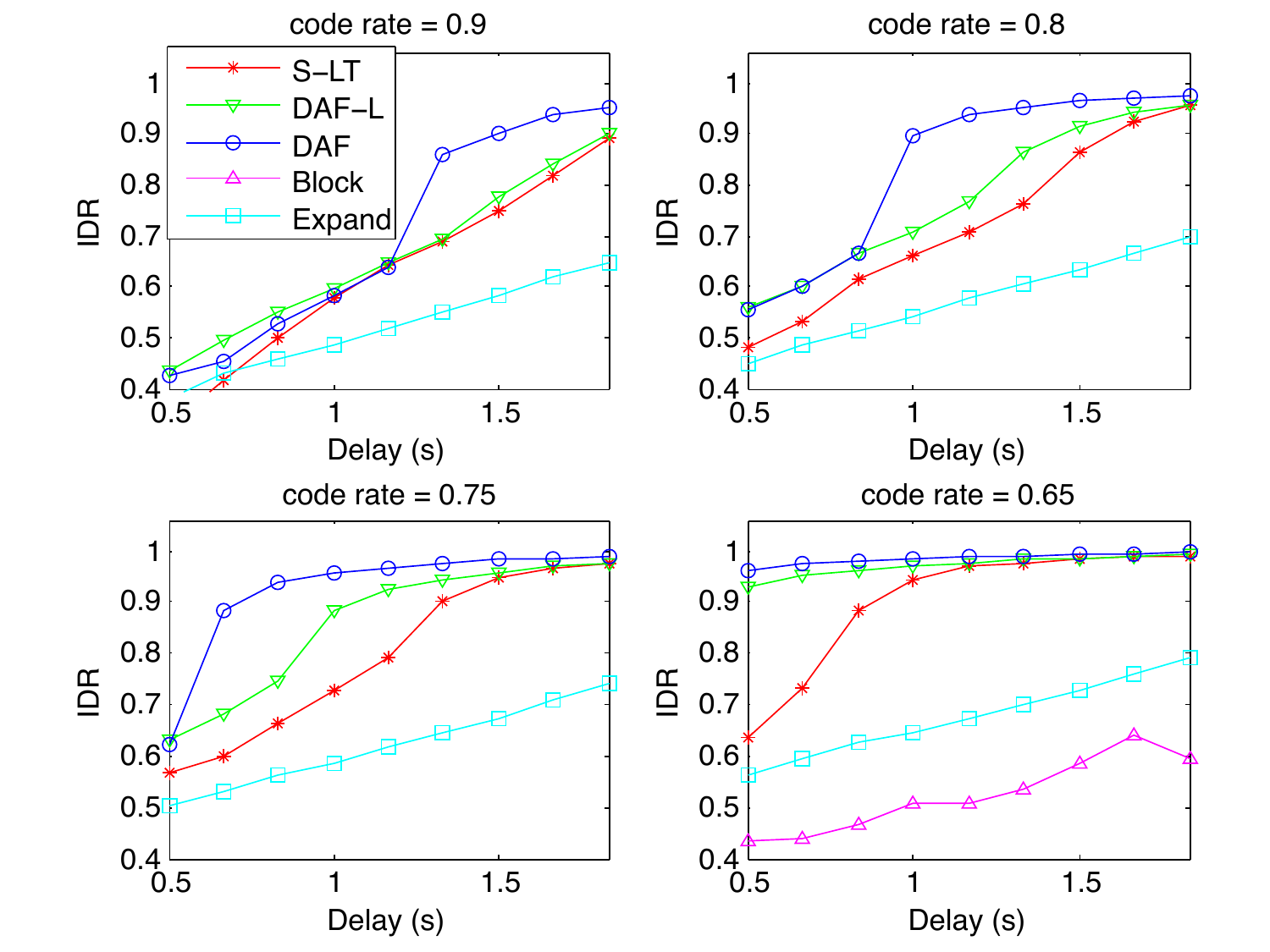}
	\caption{Relations of IDR vs. delay of CIF sequence \textit{foreman} when $C=$  $1.0$, $0.9$, $0.85$, and $0.75$. Five sliding window schemes are compared. $PLR=10\%$. $\Delta t=1$.}
	\label{fig:resultCR}
\end{figure}

We use the CIF sequence \textit{foreman} for the experiments. Fig.~\ref{fig:resultDelay} shows the relations of $IDR$ vs. $C$ of CIF sequence \textit{foreman} for different $T_{Delay}$. The results of four delays are shown: 0.5, 1, 1.5 and 1.83 seconds. Fig.~\ref{fig:resultCR} shows the relations of $IDR$ vs. $T_{Delay}$ of CIF sequence \textit{foreman} for different $C$. The results of four code rates are shown:  $1.0$, $0.9$, $0.85$, and $0.75$. Only partial results of ``block'' scheme are shown, because its values are too small to be maintained in the same scale as others. 

\begin{figure}[]
\centering
    \includegraphics[width= 0.35\textwidth]{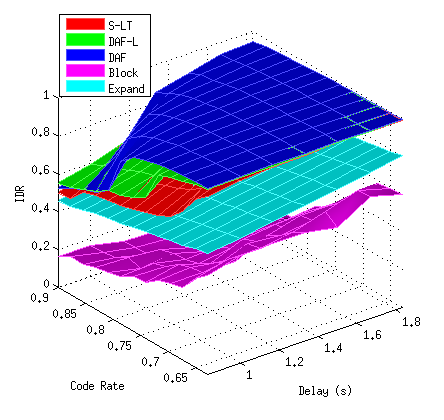}
    \caption{Comparison of IDR of \textit{foreman}. Five delay-aware fountain code schemes are compared under Case~1. }
    \label{fig:surf}
\end{figure}

We choose all the combinations of $T_{Delay} \in [0.8 , 1.8]$ and $C \in [0.6 , 0.9]$ to conduct the experiments. There are two dimensions of variables, $T_{Delay}$ and $C$, so the results of each scheme form a surface. Fig.~\ref{fig:surf} shows five surfaces of the schemes.

The numerical comparative results between different schemes with variant delays and code rates of sequences \textit{foreman} are shown in Table~\ref{tab:result-table}.

\begin{table}[] 
	\centering
	\caption{Decoding ratio comparisons under Case 1.}
	\begin{tabular}{ l|l|l|l|l|l }
		\hline
		\hline
		Code	&Data Rate	&Delay	&Scheme	&IDR	&FDR \\
		Rate	&(kbps)	&(s)	&		&		& \\
		\hline
			&	&				&DAF	&93.99\%	&98.81\%\\ 
			\cline{4-6}
			&	&				&DAF-L	&74.42\%	&98.29\%\\ 
			\cline{4-6}
			0.74		&3166	&0.8	&S-LT	&66.34\%	&82.78\%\\ 
			\cline{4-6}
			&	&				&Block	&29.32\%	&29.32\%\\ 
			\cline{4-6}
			&	&				&Expand	&56.14\%	&82.67\%\\ 
			\cline{4-6}
			&	&				&TCP	&68.21\%	&68.21\%\\ 
			\hline
			&	&				&DAF	&95.20\%	&98.72\%\\ 
			\cline{4-6}
			&	&				&DAF-L	&81.35\%	&97.94\%\\ 
			\cline{4-6}
		0.79	&3097	&1.2	&S-LT	&73.74\%	&97.26\%\\ 
			\cline{4-6}
			&	&		&Block	&27.68\%	&27.68\%\\ 
			\cline{4-6}
			&	&				&Expand	&59.21\%	&83.64\%\\ 
			\cline{4-6}
			&	&				&TCP	&62.25\%	&62.25\%\\ 
			\hline
			&	&				&DAF	&93.73\%	&98.22\%\\ 
			\cline{4-6}
			&	&				&DAF-L	&84.20\%	&96.90\%\\ 
			\cline{4-6}
			0.90		&2872	&1.7	&S-LT	&81.71\%	&96.88\%\\ 
			\cline{4-6}
			&	&				&Block	&25.14\%	&25.14\%\\ 
			\cline{4-6}
			&	&				&Expand	&61.84\%	&83.18\%\\ 
			\cline{4-6}
			&	&				&TCP	&57.23\%	&57.23\%\\ 
		\hline
		\hline
	\end{tabular}
	\label{tab:result-table}
\end{table}%

From the results above, we have the following observations:

\begin{itemize}


\item Among all schemes, DAF has the highest decoding ratio. As shown in Fig.~\ref{fig:surf}, almost the entire surface of DAF is above the other schemes. The performance of DAF-L is lower than DAF, but higher than others. DAF outperforms DAF-L because the overall sampling distribution of DAF is more homogeneous. The proposed schemes improve the decoding ratio when coding resource is insufficient or tolerable delay is small.

\item The performance of S-LT is lower than two proposed schemes, but higher than others. DAF and DAF-L outperform S-LT because their window size is bigger.

\item If $C$ is low enough or $T_{Delay}$ is large enough, the decoding ratios of all three SWFC schemes converge to 100\%. Correspondingly, if $C$ is too high or $T_{Delay}$ is too small, their performances are equally bad, or DAF may be even worse than the DAF-L scheme. That is because when data rate is extremely limited, DAF makes all the frames unlikely to be decoded at the same time, while in DAF-L scheme, some frames with very low bit rate will be decoded. However, since in those scenarios the video decoding ratios are below 50\%, which is too low to be properly viewed, they are not the cases we concern about the most.

\item The decoding ratio of all the schemes is an increasing function of $T_{Delay}$, and also a decreasing function of $C$. That means larger delay and lower code rate lead to higher overall performance, which meets our expectation. Also, Table~\ref{tab:result-table} shows that in order to obtain the decoding ratio at a certain level, we need to balance $T_{Delay}$ and $C$.


\item TCP's performance is relatively low. The reason is that TCP is not suitable for wireless scenarios where PLR is high \cite{TCPwireless}. The slow start, congestion avoidance phases and congestion control mechanisms lower its performance.

\item Block scheme performs the poorest among all schemes. Since the blocks are too small ($T_{Delay}/2$) and non-overlapping, the coding overhead is very large \cite{overhead}.

\item The above observations are true for both $IDR$ and $FDR$. There is always $FDR \ge IDR$, as we pointed out in section \ref{sec:eval-crit}. For TCP and Block schemes, there is $FDR = IDR$, because the frames prior to current window will never be decoded in the future.

\item Although decoding ratios of DAF and DAF-L are high ($90\% - 99\%$) compared to other schemes, it hardly reaches $100\%$, due to the limitations of LT code \cite{RaptorCodes}. 

\end{itemize}

\subsubsection{Case 2: Various number of hops with no node mobility}
\label{sec:case2}

The setup of this set of experiments is the following. The network consists of a source node, a destination node, and 0 or 1 or 2 relay nodes. All the nodes in the network form a chain topology from the source node to the destination node. The communication path from the source node to the destination node has 1 or 2 or 3 hops. All the nodes are immobile; hence the network topology is fixed. For all the experiments in Case 2, we set $PLR=5\%$ for each hop/link. Let $\Delta t=1$.

The $IDR$ results of sequences \textit{mobile} and \textit{akiyo} are compared in Table~\ref{tab:seq-result-table}, where N/A means the corresponding decoding ratio is below 10\% and unable to recover any consecutive frames, making the actual decoding ratio insignificant. We have the following observations:

\begin{table}[] 
	\centering
	\caption{$IDR$ comparisons under Case 2.}
	\begin{tabular}{ c|l|l|l|l|l|l }
		\hline
		\hline
					&Code		&Delay 		&		&\multicolumn{3}{ |c }{IDR}\\
		\cline{5-7}
		Sequence	&Rate	&(s)	&Scheme	&1 hop	&2 hops 		&3 hops\\
		\hline
			&	&	&DAF				&95.61\%	&90.23\%	&38.07\%\\
			\cline{4-7}
			&	&	&DAF-L				&93.22\%	&67.55\%	&36.67\%\\
			\cline{4-7}
		mobile	&0.77	&0.5	&S-LT		&66.90\%	&45.85\%	&25.97\%\\
			\cline{4-7}
			&	&	&Block		&26.91\%	&17.10\%	&14.94\%\\
			\cline{4-7}
			&	&	&Expand				&43.04\%	&40.90\%	&37.57\%\\
			\cline{4-7}
			&	&	&TCP				&82.24\%	&48.09\%	&N/A\\
			\hline
			&	&	&DAF				&94.85\%	&90.30\%	&53.18\%\\
			\cline{4-7}
			&	&	&DAF-L				&93.26\%	&82.80\%	&38.03\%\\
			\cline{4-7}
		akiyo	&0.83	&1	&S-LT		&80.68\%	&38.03\%	&30.68\%\\
			\cline{4-7}
			&	&	&Block		& N/A	& N/A	&N/A\\
			\cline{4-7}
			&	&	&Expand				&51.36\%	&49.62\%	&46.97\%\\
			\cline{4-7}
			&	&	&TCP				&76.31\%	&39.51\%	&N/A\\
		\hline
		\hline
	\end{tabular}
	\label{tab:seq-result-table}
\end{table}%

\begin{itemize}
\item The relationship of DAF, DAF-L and S-LT remains the same as in case 1 for different $PLR$: DAF achieves the highest decoding ratio among all schemes; DAF-L scheme is the second best; S-LT performs the worst among the three. That shows the proposed schemes maintain their advantages over the state-of-the-art schemes in a wide range of network conditions.

\item The performance of block coding scheme is still the lowest among all schemes.


\item TCP performs relatively well in the cases when $PLR = 5\%$, but they are extremely inefficient when $PLR=15\%$. That is because its performance is very sensitive to packet losses. High loss rate will cause TCP to time out.

\end{itemize}

\subsubsection{Case 3: Two hops with a moving relay node}
\label{sec:case3}

The setup of this set of experiments is the following. 
There are three nodes in the network: a source node and a destination node are fixed, and a relay node is moving.
The distance between the source node and the destination node is 1200 meters; the transmission range of each node is 700 meters.
Hence, the source node cannot directly communicate with the destination node; a relay node is needed.
The relay node is moving back and forth along the straight line, which is perpendicular to the straight line that links the source node and the destination node;
in addition, the relay node has equal distance to the source node and the destination node.
When the relay node moves into the transmission range of the source node, it can pick up the packets transmitted by the source node, and relay the packets to the destination node.  When the relay node moves out of the transmission range of the source node, it cannot receive packets transmitted by the source node although the source node keeps transmitting; in this case, all the packets transmitted by the source node will be lost.
The communication path from the source node to the destination node has two hops.
Since the relay node moves around, the network topology is dynamic.

In this set of experiments, we stream the sequence \textit{coastguard}, with $C = 0.8$ (the corresponding $R=2611$), and $T_{Delay} = 0.8 \text{ s}$.
Table~\ref{tab:case3} shows the $IDR$ of schemes under Case~3.
We have the following observations:

\begin{itemize}
\item DAF and DAF-L still perform the best among all the schemes. However, the decoding ratios are not as high as previous cases. That is because when the relay node temporarily moves out of the connection range, the source does not stop streaming video, therefore the content transmitted during the disconnecting period is lost.

\item The performance of S-LT is worse than proposed schemes but better than others, and the block coding scheme is still among the worst schemes, for the same reasons as in Case~1.


\item The performance of TCP scheme is also poor, because the disconnecting period causes the time-out in TCP. 
\end{itemize}

\begin{table}[] 
	\centering
	\caption{$IDR$ comparisons under Case 3. }
	\begin{tabular}{ l|l }
		\hline
		\hline
		Scheme	&IDR \\
		\hline
			DAF	&85.16\%\\
			\hline
			DAF-L	&65.21\%\\
			\hline
			S-LT	&32.42\%\\
			\hline
			Block	&N/A\\
			\hline
			Expand	&20.90\%\\
			\hline
			TCP	&N/A\\
		\hline
		\hline
	\end{tabular}
	\label{tab:case3}
\end{table}%

\section{Conclusions}
\label{sec:conclusions}

This paper proposed a novel delay-aware fountain code scheme for video streaming, which deeply integrates channel coding and video coding. This is the first work to exploit the fluctuation of bit rate in video data at the level of channel coding, and to incorporate it towards the optimal design of video streaming-oriented fountain codes. Specifically, we developed two novel coding techniques: the time-based sliding window and the optimal window-wise sampling strategy. The proposed scheme delivers significantly higher video quality than existing schemes with a constant bandwidth cost. We designed two protocols, DAF and DAF-L, to improve video decoding ratio under different computational complexity. The simulation results show that the decoding ratio of our scheme is 15\% to 100\% higher than the state-of-the-art delay-aware schemes in a variety of settings.


\bibliographystyle{IEEEtran}
\bibliography{IEEEabrv,delay}

\end{document}